\documentclass[12pt]{article}

\begin{document}

\begin{center}
{\bf Equations for massless and massive spin-1/2 particles with varying speed and neutrino in matter} \\
\vspace{5mm} S. I. Kruglov
\footnote{E-mail: serguei.krouglov@utoronto.ca}

\vspace{3mm}
\textit{Department of Chemical and Physical Sciences, University of Toronto,\\
3359 Mississauga Rd. North, Mississauga, Ontario, Canada L5L 1C6} \\
\vspace{5mm}
\end{center}

\begin{abstract}
The modified Dirac equations describing massless and massive spin-1/2 particles violating the Lorentz invariance are considered. The equation for massless fermions with varying speed is formulated in the 16-component first-order form. The projection matrix, which is the density matrix, extracting solutions to the equation has been obtained. Exact solutions to the equation and energy spectrum for a massive neutrino are obtained in the presence of background matter. We have considered as subluminal as well as superluminal propagations of neutrinos. The bounds on the Lorentz invariance violation parameter from astrophysical observations by IceCube, OPERA, MINOS collaborations and by the SN1987A supernova are obtained for superluminal neutrinos.
\end{abstract}

\section{Introduction}

It follows from the string theory that the Lorentz and CPT symmetries can be broken \cite{Samuel}, \cite{Colladay1}.
The effective field theory \cite{Colladay}, \cite{Coleman} describes the Lorentz invariance violation (LIV) and gives bounds on LIV coefficients \cite{Russel}. There are investigations of LIV of the photon field \cite{Carroll}, \cite{Kruglov0}, \cite{Kostelecky} and the fermion field \cite{Ferreira}, \cite{Bolokhov}, \cite{Kostelecky1}, \cite{Kostelecky2}. It should be mentioned that an extensive description of Lorentz-violating fermions is presented in \cite{Kostelecky0}. Effects of LIV are of the order of the Planck mass $M_{Pl}=1.22\times 10^{19}$ GeV and therefore can be found at high energies. Possibly LIV is due to quantum gravity corrections \cite{Amelino}, \cite{Smolin} and can lead to deformed dispersion relations:
\begin{equation}
p_0^2=\textbf{p}^2+m^2-\left(Lp_0\right)^\alpha\textbf{p}^2,
\label{1}
\end{equation}
where $p_0$ is an energy and the LIV parameter $L$ possesses the dimension of ``length". We suppose that $L>0$ and $L$ is of the order of the Plank length $L_{Pl}=M_{Pl}^{-1}$. It should be mentioned that space-time foam Liouville-string models \cite{Ellis} also lead to the modified dispersion relation (1) with the parameter $\alpha=1$. The wave equation for massive spinless particles within the dispersion equation (1) at $\alpha=1$ was considered in \cite{Ellis2}, \cite{Kruglov} and for massless particles in \cite{Kruglov1}. It should be mentioned that firstly the wave equation for spinless particles with the breakdown of Lorentz invariance was considered in \cite{Pavlopoulos}. The wave equation for massive fermions in the framework of the dispersion equation (1) at $\alpha=2$ was introduced in \cite{Kruglov2}. Here we consider the equation and its solutions for massless particles with spin-1/2 within the deformed dispersion relation (1) at $\alpha=2$. Also we obtain exact solutions to the equation for a massive spin-1/2 particle in the case of particle-matter interaction.

In Sec.2 we introduce the wave equation for massless and massive particles with LIV parameter realizing the dispersion relation (1) with $\alpha=2$. The 16-component first-order form of equations is obtained. The solutions to the wave equation for massless particles in the form of the projection matrix are found in Sec.3. In Sec.4 we consider the massive fermion particles (neutrinos) in the presence of background matter. The exact solutions to the equation with particle-matter interaction are obtained. In Sec.5 we consider superluminal neutrinos by modification of the wave equation and the dispersion relation. The bounds on the Lorentz invariance violation parameter from astrophysical observations by IceCube, OPERA, MINOS collaborations and by the SN1987A supernova are obtained. We discuss the results obtained in Sec.6. In the Appendix some products of the $16$-component matrices, which enter the wave equations, are found.

We use the system of units $\hbar =c=1$, Greek letters run 1,2,3,4 and Latin letters run 1,2,3.

\section{Modified equations for spin-1/2 particles}

Let us introduce the system of first-order equations for spin-1/2 particles:
\[
\gamma_\mu\partial_{\mu}\psi(x)+m_0-im_1L\gamma_4\partial_t\gamma_i\psi_i(x)=0,
\]
\vspace{-7mm}
\begin{equation} \label{2}
\end{equation}
\vspace{-7mm}
\[
m_1\psi_i(x)=\partial_i\psi(x),
\]
where $\gamma_\mu$ are Dirac's matrices, $\partial_\mu=\partial/\partial x_\mu=\left(\partial/\partial x_i,\partial/(i\partial t)\right)$, $x_0=t$ is a time; $m_0$ and $m_1$ are parameters with the dimension of the mass, $\psi(x)$ is bispinor, and we use the notations of \cite{Ahieser}.
The mass parameter $m_1$ is introduced to have the same dimension of all terms in Eqs.(2). The physical values calculated should not depend on $m_1$. Eqs.(2) are covariant under the group of rotations but the invariance under the boost transformations is violated due to the additional LIV term containing the parameter $L$. Such term corresponds to the isotropic LIV. As a result, the Lorentz symmetry is broken and one needs to introduce preferred frame of reference. It should be noted that the isotropic Lorentz invariance violation for fermion fields was discussed in \cite{Kostelecky1}.
We treat Eqs.(2) as effective wave equations with preferred frame effects. From (2) one obtains the equation for massive fermions with LIV suggested in \cite{Kruglov2}:
\begin{equation}
\gamma_\mu\partial_{\mu}\psi(x)+m_0-iL\gamma_4\partial_t\gamma_i\partial_i\psi(x)=0.
\label{3}
\end{equation}
The dispersion equation (1) at $\alpha=2$ follows from Eq.(3) at $m_0=m$ \cite{Kruglov2}. At $m_0=0$ we find from Eqs.(2) the equation for massless spin-1/2 particles
\begin{equation}
\gamma_\mu\partial_{\mu}\psi(x)-iL\gamma_4\partial_t\gamma_i\partial_i\psi(x)=0
\label{4}
\end{equation}
realizing the dispersion relation (1) at $m=0$, $\alpha=2$. Such dispersion relation within Eq.(4) leads to the speed of massless fermions depending on the energy $p_0$ and corresponds to subluminal particle propagation.

\subsection{Wave equation for particles in the first-order 16-component form}

Following \cite{Kruglov2} we define the 16-component wave function
\begin{equation}
\Psi (x)=\left\{ \Psi _A(x)\right\} =\left(
\begin{array}{c}
\psi(x)\\
\psi_i (x)
\end{array}
\right),
\label{5}
\end{equation}
using the notations  $\Psi_0(x)=\psi(x)$, $\Psi_i(x)=(1/m_1)\partial_i\psi(x)$. With the help of the elements of the entire matrix algebra $\varepsilon^{A,B}$, having matrix elements and products of matrices \cite{Kruglov3}
\begin{equation}
\left( \varepsilon ^{M,N}\right) _{AB}=\delta _{MA}\delta _{NB},
\hspace{0.5in}\varepsilon ^{M,A}\varepsilon ^{B,N}=\delta
_{AB}\varepsilon ^{M,N},
\label{6}
\end{equation}
where $A,B,M,N=(0,i)$, the system of equations (2), with Eq.(5), can be written in the 16-component matrix form
\[
\biggl[\left(\varepsilon ^{0,0}\otimes\gamma_\mu+ m_1L\delta_{4\mu}\varepsilon
^{0,i}\otimes\gamma_4\gamma_i-\delta_{\mu i}\varepsilon ^{i,0}\otimes I_4\right)\partial_\mu
\]
\vspace{-8mm}
\begin{equation}
\label{7}
\end{equation}
\vspace{-8mm}
\[
+ \left(m_0\varepsilon ^{0,0}+m_1\varepsilon ^{i,i}\right)\otimes I_4
\biggr]\Psi(x)=0 ,
\]
where $I_4$ being the $4\times 4$ identity matrix, the $\otimes$ is the direct product of matrices. One implies the summation over all repeated indices.  Let us introduce the
$16\times 16$-matrices
\[
\Gamma_\mu=\varepsilon ^{0,0}\otimes\gamma_\mu+ m_1L\delta_{4\mu}\varepsilon
^{0,i}\otimes\gamma_4\gamma_i-\delta_{\mu i}\varepsilon ^{i,0}\otimes I_4,
\]
\vspace{-7mm}
\begin{equation} \label{8}
\end{equation}
\vspace{-7mm}
\[
P_0=\varepsilon ^{0,0}\otimes I_4,~~~~P_1=\varepsilon ^{i,i}\otimes I_4,
\]
where $P_0$, $P_1$ are projection matrices, $P_0^2=P_0$, $P_1^2=P_1$. Then Eq.(7) can be represented in the matrix first-order $16\times16$-component form
\begin{equation}
\left( \Gamma_\mu \partial _\mu + m_0P_0+m_1P_1\right)\Psi(x)=0 .
\label{9}
\end{equation}
The generalized Dirac matrices are given by
\begin{equation}
\Gamma_m=\varepsilon ^{0,0}\otimes\gamma_m-\varepsilon ^{m,0}\otimes I_4,~~~
\Gamma_4=\varepsilon ^{0,0}\otimes\gamma_4+ m_1L\varepsilon
^{0,i}\otimes\gamma_4\gamma_i.
\label{10}
\end{equation}
It is convenient for massive particles to put $m_0=m_1=m$. Then Eq.(9) is simplified and we come to the equation
\begin{equation}
\left( \Gamma_\mu \partial _\mu + m\right)\Psi(x)=0 ,
\label{11}
\end{equation}
as $P_0+P_1=I_{16}$, where $I_{16}$ is the unit $16\times16$-matrix. Eq.(11) was considered in \cite{Kruglov2}. For the case of massless fermions we put $m_0=0$ in Eq.(9). As a result, we come to the $16\times16$-matrix equation for massless spin-1/2 particles
\begin{equation}
\left( \Gamma_\mu \partial _\mu + m_1P_1\right)\Psi(x)=0 .
\label{12}
\end{equation}
Solutions of Eq.(11) in the form of projection operators were found in \cite{Kruglov2}. Now we obtain solutions to Eq.(12) describing massless fermions.

\section{Solutions to the wave equation for massless particles}

The wave function for the positive energies becomes $\Psi(x)=\Psi(p) \exp[i(\textbf{p}\textbf{x}-p_0x_0)]$ and Eq.(12) reads
\begin{equation}
\left( i\check{p} + m_1P_1\right) \Psi (p)=0 , \label{13}
\end{equation}
where $\check{p}=\Gamma_\mu p _\mu$. The matrix $\check{p}$ obeys the equation as follows \cite{Kruglov2}
\begin{equation}
\check{p}^5 -p^2\check{p}^3+\left(m_1L\right)^2p^2_4\textbf{p}^2\check{p}=0 , \label{14}
\end{equation}
where $p_4=ip_0$.
Let us introduce the matrix of Eq.(13)
\begin{equation}
\Omega = i\check{p} + m_1P_1.
\label{15}
\end{equation}
With the help of Eqs.(6),(15), and taking into account dispersion equation (1) for $m=0$, $\alpha=2$, one can obtain the matrix equation (see Appendix)
\begin{equation}
\Omega\left(\Omega-m_1\right)\left[\Omega\left(\Omega-m_1\right)^2+p^2\left(\Omega-2m_1\right)\right]=0.
\label{16}
\end{equation}
Using Eq.(16), we find solutions to Eq.(13) in the form of the projection matrix
\begin{equation}
\Xi=N\left(\Omega-m_1\right)\left[\Omega\left(\Omega-m_1\right)^2+p^2\left(\Omega-2m_1\right)\right],
 \label{17}
\end{equation}
where $N$ is the normalization constant, and $\left(i\check{p} + m_1P_1\right)\Xi=0$. As a result, every column of the matrix $\Xi$ is the solution to Eq.(13). The constant $N$ can be obtained from the requirement \cite{Fedorov}:
\begin{equation}
\Xi^2=\Xi. \label{18}
\end{equation}
We find from Eqs.(17),(18), after some calculations, the normalization constant
\begin{equation}
N=\frac{1}{2m_1^2p^2}.
\label{19}
\end{equation}
The matrix $\Xi$ is a density matrix for impure spin states, and to consider pure spin states, we introduce the spin projection operator \cite{Kruglov5}
\begin{equation}
\sigma_p=-\frac{i}{2|\textbf{p}|}\epsilon_{abc}\textbf{p}_a
J_{bc},
\label{20}
\end{equation}
where $|\textbf{p}| =\sqrt{p_i^2}$, and
\begin{equation}
J_{mn}=\left(\varepsilon^{m,n}- \varepsilon^{n,m}\right)\otimes I_4
+I_4\otimes \frac{1}{4}\left(\gamma_m\gamma_n-\gamma_n\gamma_m\right).
\label{21}
\end{equation}
are generators of the rotational group in the $16$-dimension representation space. The projection operators extracting spin projections $\pm 1/2$ are given by \cite{Kruglov5}
\begin{equation}
P_{\pm 1/2}=\mp\frac{1}{2}\left(\sigma_p\pm\frac{1}{2}
\right)\left(\sigma_p^2-\frac{9}{4}\right).
 \label{22}
\end{equation}
The projection operator which extracts the states with positive energy and definite spin projection is given by
\begin{equation}
\Delta^{(0)}_{\pm 1/2}= \Xi P_{\pm 1/2}, \label{23}
\end{equation}
and obeys the matrix equation $\Delta^{(0)2}_{\pm 1/2}=\Delta^{(0)}_{\pm 1/2}$.
The projection operator $\Delta^{(0)}_{\pm 1/2}$ can be used for calculations of quantum processes with massless spin-1/2 fermions.

\section{Neutrinos in the presence of background matter}

Now we apply Eq.(3) (with $m_0=m$) for a description of the propagation of a massive neutrino through background dense matter. The presence of LIV can generate the definite effects in astrophysical environments.
We imply that neutrinos possess small masses and interact weakly with the matter. When neutrinos propagate through neutron stars electro-weak interaction should be taken into consideration.
To introduce the interaction of neutrino with the matter we follow the works of \cite{Studenikin1} - \cite{Studenikin4}. Thus, one introduces the additional term in Eq.(3) due to the interaction of neutrino with the unpolarized and stationary background matter
\begin{equation}
\left[\gamma_\mu\partial_{\mu}+m+\frac{f_4}{2}\gamma_4\left(1+\gamma_5\right)
-iL\gamma_4\partial_t\gamma_i\partial_i\right]\psi(x)=0,
\label{24}
\end{equation}
where $f_4=(G_F/\sqrt{2})\rho_n$, $G_F$ is the Fermi constant and $\rho_n$ being the neutron number density (in a
typical neutron star $\rho_n\approx 0.4$ fm$^{-3}$ and $G_F\rho_n/\sqrt{2}\sim 20$ eV \cite{Kiers}). The additional term introduced represents the electro-weak potential due to the finite neutron density. We use matrices as follows \cite{Ahieser}:
\[
\alpha_k=i\gamma_4\gamma_k=\left(
\begin{array}{cc}
0&\sigma_k\\
\sigma_k& 0
\end{array}
\right)
,~~~
\gamma_4=\left(
\begin{array}{cc}
I_2&0\\
0&-I_2
\end{array}
\right),
\]
\vspace{-7mm}
\begin{equation} \label{25}
\end{equation}
\vspace{-7mm}
\[
\gamma_5=\gamma_1\gamma_2\gamma_3\gamma_4=\left(
\begin{array}{cc}
0&-I_2\\
-I_2&0
\end{array}
\right),
\]
where $\sigma_k$ are Pauli matrices and $I_2$ is $2\times 2$ identity matrix.
Using the bispinor wave function
\begin{equation}
\Psi\left(x\right)=\left(
\begin{array}{c}
\varphi (x)\\
\chi (x)
\end{array}
\right),
\label{26}
\end{equation}
for the positive energy $\psi(x)=\psi(p) \exp[i(\textbf{p}\textbf{x}-p_0x_0)]$, with the help of Eqs.(25), we obtain from Eq.(24) the system of equations
\[
\left(1-Lp_0\right)p_k\sigma_k \chi(p)+\frac{f_4}{2}\left[\varphi(p)-\chi(p)\right]
=\left(p_0-m\right)\varphi(p),
\]
\vspace{-8mm}
\begin{equation}
\label{27}
\end{equation}
\vspace{-8mm}
\[
-\left(1+Lp_0\right)p_k\sigma_k\varphi(p) +\frac{f_4}{2}\left[\varphi(p)-\chi(p)\right]
=-\left(p_0+m\right)\chi(p).
\]
From Eqs.(27) one finds the equation for spinor $\varphi(p)$:
\begin{equation}
\left[\left(1-L^2p^2_0\right)\textbf{p}^2
-f_4p_k\sigma_k+f_4p_0+m^2-p_0^2\right]\varphi(x)=0.
\label{28}
\end{equation}
We take into account the helicity of neutrino, and come to equations
\begin{equation}
p_k\sigma_k\varphi(p)=s|\textbf{p}|\varphi(p),~~~p_k\sigma_k\chi(p)=s|\textbf{p}|\chi(p),
\label{29}
\end{equation}
where $s=\pm 1$ is the helicity of neutrino. Then Eq.(28) leads to the energy spectrum
\begin{equation}
p_0^2= m^2+\left(1-L^2p_0^2\right)\textbf{p}^2+f_4\left(p_0-s|\textbf{p}|\right).
\label{30}
\end{equation}
At $f_4=0$ we arrive to the dispersion equation (1) ($\alpha=2$). Thus, Eq.(30) takes into consideration both the effect of LIV containing the parameter $L$ and the interaction of neutrino with the matter.
The quadratic equation (30) has the solution for the positive energy as follows
\begin{equation}
p_0=\frac{\tilde{n}}{1+L^2\textbf{p}^2} +\frac{1}{\sqrt{1+L^2\textbf{p}^2}}\sqrt{m^2+\textbf{p}^2-2s\tilde{n}|\textbf{p}|+\frac{\tilde{n}^2}{1+L^2\textbf{p}^2}},
\label{31}
\end{equation}
where $\tilde{n}=f_4/2$. In the order of ${\cal O}(L^2\textbf{p}^2)$, ${\cal O}(\tilde{n}^2)$, we obtain from Eq.(31) the approximate equation
\begin{equation}
p_0\approx\tilde{n}+\sqrt{m^2+\textbf{p}^2-2s\tilde{n}|\textbf{p}|}\left(1-\frac{1}{2}L^2\textbf{p}^2\right) .
\label{32}
\end{equation}
At $L=0$ one finds from Eq.(32) the energy spectrum obtained in \cite{Studenikin1}.\\
From Eqs.(29) with the help of Pauli matrices we obtain the spinor
\begin{equation}
\varphi(p)=N_1\left(
\begin{array}{c}
1\\
\frac{p_1+ip_2}{p_3+s|\textbf{p}|}
\end{array}
\right),
\label{33}
\end{equation}
where $N_1$ is the normalization constant. Introducing the phase, $\tan\delta=p_2/p_1$, and changing the normalization constant, spinor $\varphi(p)$ can be represented as
\begin{equation}
\varphi(p)=N\left(
\begin{array}{c}
\sqrt{1+s\frac{p_3}{|\textbf{p}|}}\\
\sqrt{1-s\frac{p_3}{|\textbf{p}|}}\exp(i\delta)
\end{array}
\right).
\label{34}
\end{equation}
From Eqs.(27) we find the spinor $\chi(p)$ entering the wave function (26)
\begin{equation}
\chi(p)=\frac{(1+Lp_0)s|\textbf{p}|-\tilde{n}}{p_0+m-\tilde{n}}\varphi(p).
\label{35}
\end{equation}
To find the normalization constant $N$ one can use the equation as follows \cite{Ahieser}, \cite{Sokolov}:
\begin{equation}
\int\Psi^+(x)\Psi(x)dx =1.
\label{36}
\end{equation}
where $\Psi^+(x)$ is the Hermitian conjugate function. At $L=0$ from Eqs.(34),(35) we come to the case considered in \cite{Studenikin3}. If the LIV parameter $L$ is much greater than the Planck length $L_P$, then the LIV effect should be taken into account. The exact solutions (34),(35) obtained allow us to study different processes of neutrino interactions by taking into account quantum gravity corrections with LIV parameter $L$ and background matter.

It has to be noted that according to Eq.(32) the neutrino energy depends on helicity state. One can see from Eqs.(31), (32) that the energy is greater for $s=-1$ and less for $s=1$. Therefore, neutrino can emit photon by changing the helicity states. For such a proses the conservation of energy and momentum read (see \cite{Studenikin1}) \begin{equation}
(p_0)_i=(p_0)_f+\omega,~~~\textbf{p}_i=\textbf{p}_f+\textbf{k}.
\label{37}
\end{equation}
Then implying that $Lp_0\ll 1$, $\tilde{n}|\textbf{p}|/p_0^2\ll 1$, one finds from Eq.(32), in the linear approximation, the neutrino energy
\begin{equation}
p_0\approx E_0+\tilde{n}-\frac{s\tilde{n}|\textbf{p}|}{E_0}-\frac{1}{2}L^2\textbf{p}^2E_0,
\label{38}
\end{equation}
where $E_0=\sqrt{m^2+\textbf{p}^2}$. From the conservation of energy and momentum, Eqs.(37), the energy of the emitted photon was obtained in \cite{Studenikin2}, \cite{Studenikin3}
\begin{equation}
\omega=(s_f-s_i)\tilde{n}\frac{\beta}{1-\beta \cos \theta},
\label{39}
\end{equation}
where $\theta$ is the angle between the directions of neutrino and photon propagations, $\cos \theta=(\textbf{p}_i\cdot \textbf{k})/(|\textbf{p}_i||\textbf{k}|)$, $\beta=|\textbf{p}|/E_0$ is the speed of neutrino in the absence of LIV parameter $L$. We obtain from Eq.(30) for ultra-relativistic neutrino in the linear approximation ($m\ll p_0$):
\begin{equation}
\frac{|\textbf{p}|}{p_0}=1-\frac{m^2}{2p^2_0}+\frac{(s-1)\tilde{n}}{p_0}+\frac{L^2p_0^2}{2}.
\label{40}
\end{equation}
It should be mentioned that the neutrino speed is defined by the group velocity $v=\partial p_0/\partial |\textbf{p}|$ and it is different from $\beta$ due to dispersion relation (30). Neglecting the LIV parameter, one has $\beta =v$. But in the background matter and taking into account quantum gravity corrections, for ultra-high energy in the linear approximation, we obtain from Eqs.(30),(40) the neutrino speed
\begin{equation}
v=1-\frac{m^2}{2p^2_0}-\frac{3}{2}L^2p_0^2.
\label{41}
\end{equation}
The last term in Eq.(41)) is due to quantum gravity corrections and gives the contribution to the speed of neutrinos.
The phenomenon of the radiation of photons by neutrino propagating in the matter was named in \cite{Studenikin1} the spin light, $SL\nu$. Thus, the LIV term in the Dirac equation introduced \cite{Kruglov2} contributes to $SL\nu$ for high energetic neutrinos.

It should be mentioned that according to Eq.(38) $\delta p_0=p_0-E_0<0$ (for a case of the propagation of a neutrino without the interaction with the matter, $\tilde{n}=0$), and there are no bounds on LIV parameter $L$ from hadron threshold effects, \v{C}erenkov-like decays and deviation of the neutrino speed within IceCube \cite{IceCube}, \cite{IceCube1}, \cite{IceCube2}, OPERA \cite{OPERA}, MINOS \cite{MINOS} and SN1987A supernova \cite{Hirata} observations. This is a consequence of a subluminal propagation of neutrinos in the model under consideration. The corresponding bounds on LIV parameters were made in the general approach of works \cite{Kostelecky1}, \cite{Kostelecky2}, \cite{Kostelecky0}. There are also the stringent constraints on LIV from the two IceCube PeV neutrinos in \cite{Chakraborty}, \cite{Stecker} for the case of superluminal neutrinos.

\section{Superluminal neutrinos}

To describe superluminal neutrinos we should modify the wave equation (24) and dispersion relation (1). With the help of the replacement $L=i\Lambda$, with the real value of $\Lambda$, we obtain from Eq.(1) the dispersion relation for the case of $\alpha=2$
\begin{equation}
p_0^2=\textbf{p}^2+m^2+\left(\Lambda p_0\right)^2\textbf{p}^2.
\label{42}
\end{equation}
Then the group velocity (41) for ultra relativistic neutrinos becomes
\begin{equation}
v=1-\frac{m^2}{2p^2_0}+\frac{3}{2}\Lambda^2p_0^2.
\label{43}
\end{equation}
At $1.5\Lambda^2p_0^2>m^2/(2p^2_0)$, we have superluminal propagation of neutrinos, $v>1$. The wave equation for massive neutrinos (without the interaction with the matter) leading to the dispersion relation (42) follows from Eq.(24)
\begin{equation}
\left[\gamma_\mu\partial_{\mu}+m +\Lambda\gamma_4\partial_t\gamma_i\partial_i\right]\psi(x)=0.
\label{44}
\end{equation}
Now, we consider bounds on the parameter $\Lambda$ in the case of superluminal propagation of neutrinos from IceCube, OPERA, MINOS collaborations and by the SN1987A supernova observations.

\subsection{Bounds from hadron threshold effects and \v{C}erenkov-like decays}

Threshold effects in the pion decay can be used to obtain a constraint on the LIV parameter $\Lambda$. The energy-momentum conservation for the decay process $\pi^+\rightarrow \mu^++\nu_\mu$ gives the restriction \cite{Kostelecky1}
\begin{equation}
\delta p_0\leq \frac{\Delta M^2}{2|\textbf{p}|},
\label{45}
\end{equation}
where $\delta p_0$ is the Lorentz-violating contribution to the energy, $\Delta M=M_\pi -M_\mu$ being the mass difference of a pion and a muon. From Eq.(38) we obtain
\begin{equation}
\delta p_0=\frac{1}{2}\Lambda^2\textbf{p}^2E_0.
\label{46}
\end{equation}
Neglecting the contribution of the matter interaction, one finds from Eqs.(45), (46)
\begin{equation}
\Lambda^2|\textbf{p}|^4\leq \Delta M^2,
\label{47}
\end{equation}
where we use the approximation for high energy neutrinos $p_0\approx E_0\approx |\textbf{p}|$.
For the pion decay we have $\Delta M^2=1.14\times10^{-3}$ $\mbox {GeV}^2$. It should be noted that
the bound for $\Lambda$ is less for the kaon decay and can be ignored.
At high energies the IceCube collaboration observed the energy of atmospheric neutrinos up to $400$ TeV \cite{IceCube} and two PeV events \cite{IceCube1}, \cite{IceCube2}. Thus, with the help of IceCube collaboration data ($|\textbf{p}|=400$ TeV) we obtain from Eq.(47) the bound on the LIV parameter
\begin{equation}
\Lambda\leq 2.1\times 10^{-13}~\mbox {GeV}^{-1}.
\label{48}
\end{equation}
There are also the stringent constraints on LIV from the two IceCube PeV neutrinos in \cite{Chakraborty}, \cite{Stecker} for the case of superluminal neutrinos.
It should be mentioned that Planck's length $L_{Pl}=1/M_{Pl}=8.2\times10^{-20}~\mbox {GeV}^{-1}$ is much less than the upper bound of $\Lambda$. As there is no significant difference in the total neutrino
and antineutrino flux, the constraint (48) can be used for muon neutrinos and antineutrinos. At higher energies the decays of short-lived charmed hadrons with $M=2$ GeV dominate \cite{Kostelecky2}. Here we suppose that muon neutrinos or antineutrinos are due to the atmospheric hadron decay $h\rightarrow \mu+\nu_\mu$.
Thus, with the help of Eq.(47) and the value $\Delta M^2\approx 4~\mbox {GeV}^2$, where $\Delta M=M_h-M_\mu$, for the two PeV events of atmospheric neutrinos ($|\textbf{p}|=1$ PeV), one obtains the bound $\Lambda\leq 2\times 10^{-12}~\mbox {GeV}^{-1}$ which is weaker compared with (48).\\
Let us suppose that the high-energy neutrinos detected by IceCube are of astrophysical origin. For \v{C}erenkov-like decays, neutrinos have energies near the threshold and the relation $|\textbf{p}|\delta p_0\approx 2m_e^2$ ($m_e$ is the electron mass) holds \cite{Kostelecky2}. Then from Eq.(46), one has $\Lambda|\textbf{p}|^2\leq 2m_e$. For $|\textbf{p}|=1$ PeV neutrinos we obtain the stringent bound $\Lambda\leq 10^{-15}~\mbox {GeV}^{-1}$.

\subsection{Bounds from the deviation of neutrino speed}

A difference between the speed of light and the speed of muon neutrinos was reported by the OPERA neutrino experiment at the underground Gran Sasso Laboratory \cite{OPERA}, $\delta v =[ 2.7 \pm 3.1(stat.)^{+3.4}_{-3.3}(sys.)]\times 10^{-6}$. From Eq.(43) one finds
\begin{equation}
\delta v=-\frac{m^2}{2p^2_0}+\frac{3}{2}\Lambda^2p_0^2.
\label{49}
\end{equation}
Assuming that the neutrino mass is as much as $m=2$ eV and the energy of neutrinos $\langle p_ 0\rangle=17$ GeV, we find the correction to the speed of neutrino due to the neutrino mass, $m^2/(2p^2_0)\approx 7\times 10^{-20}$, which is very small. Therefore, we can suppose that the main deviation of neutrino speed from the speed of light is due to the LIV parameter $\delta v\approx 1.5\Lambda^2p_0^2$. Then we obtain the bound for $17$ GeV energetic neutrinos with the deviation of the speed $\delta v = 2.7 \times 10^{-6}$ (but the accuracy of this value is not good enough)
\begin{equation}
\Lambda\leq 7.9\times 10^{-5}~\mbox {GeV}^{-1}.
\label{50}
\end{equation}
The constraint (50) is weaker than the cosmic-ray bound (48).

The MINOS Collaboration \cite{MINOS} reported a difference in the neutrino speed $\delta v=(5.1\pm 2.9)\times 10^{-5}$ for the energy $p_0=3$ GeV. Then from Eq.(46), one finds the bound $\Lambda\leq 1.9\times 10^{-3}~\mbox {GeV}^{-1}$.

At lower energy of $10$ MeV, a limit $\delta v < 2 \times 10^{-9}$
was reported by the observation of (anti)neutrinos emitted by the SN1987A supernova \cite{Hirata}. From Eq.(46) we obtain the weak bound  $\Lambda<3.7\times 10^{-3}~\mbox {GeV}^{-1}$. As a result, the bounds from the deviation of speed of superluminal neutrinos reported by OPERA, MINOS collaborations and by the SN1987A supernova observations are weaker than bounds obtained with the help of IceCube collaboration data.

\section{Conclusion}

We have considered the wave equations for massless and massive spin-1/2 particles which lead to the modified dispersion relation discussed in the framework of quantum gravity. The solutions of 16-component first-order equations in the form of projection operators obtained can be used in different calculations of quantum processes. The projection matrix found represents the density matrix by taking into account LIV effects. We have obtained exact solutions to the wave equation for neutrino in the field of background matter. Exact solutions obtained can be used for the analysis of the propagation of neutrinos with taking into consideration LIV corrections. We also have considered the modified wave equation and dispersion relation for superluminal neutrinos. In this case
we have found the bounds on the LIV parameter $\Lambda$ from astrophysical observations by IceCube, OPERA, MINOS collaborations and by the SN1987A supernova from the threshold effect for hadrons, \v{C}erenkov-like decays and the deviation of the speed of high-energy neutrinos. The strongest bound $\Lambda\leq 10^{-15}~\mbox {GeV}^{-1}$ was obtained for $1$ PeV neutrinos observed by IceCube collaboration. Thus, the approach suggested allows us to study as subluminal as well as superluminal propagations of neutrinos. The further experimental investigations of neutrino propagations with higher accuracy will help to understand physics beyond the Standard Model.

\vspace{5mm}
\textbf{Appendix}
\vspace{5mm}

From Eqs.(10), (15) with the aid of Eqs.(6) we obtain the products of matrices
\begin{equation}
\Omega\left(\Omega-m_1\right)=-\check{p}^2-im_1\varepsilon^{0,0}\otimes \hat{p},
\label{51}
\end{equation}
\begin{equation}
\left(\Omega^2+p^2\right)\left(\Omega-m_1\right)^2=\check{p}^4-p^2\check{p}^2,
\label{52}
\end{equation}
\begin{equation}
\left(\Omega-m_1\right)\left[\Omega^2\left(\Omega-m_1\right)^2+p^2\Omega\left(\Omega-2m_1\right)+p^2m_1^2
-\left(Lp_0\right)^2m_1^2\textbf{p}^2\right]=0,
\label{53}
\end{equation}
where $\hat{p}=p_\mu \gamma_\mu$, $p^2= \textbf{p}^2-p_0^2$. With the help of dispersion relation (1) we obtain (on-shell) the minimal polynomial
\begin{equation}
\Omega\left(\Omega-m_1\right)\left[\Omega\left(\Omega-m_1\right)^2+p^2\left(\Omega-2m_1\right)\right]=0.
\label{54}
\end{equation}

\end{document}